\documentclass{icrc2009}

\usepackage{graphicx}   
\usepackage{caption}    
\usepackage[font=footnotesize]{subfig} 
\usepackage{fixltx2e}
\usepackage{url}

\newcommand{\shorttitle}[1]%
{\markboth{Proceedings of the 31\MakeLowercase{$^{st}$} ICRC, {\L}\'{o}d\'{z} 2009}{#1} }
\newcommand{\etal}{\MakeLowercase{\textit{et al. }}} 

\begin{document}
\title{Point source searches with the ANTARES neutrino telescope}
\author{\IEEEauthorblockN{Simona Toscano\IEEEauthorrefmark{1}, for the ANTARES Collaboration}

                            \\
\IEEEauthorblockA{\IEEEauthorrefmark{1} IFIC- Instituto de F\'isica
                            Corpuscular, Edificios de Investigaci\'on de
                            Paterna, CSIC - Universitat de Val\`encia,\\
                            Apdo. de Correos 22085, 46071 Valencia, Spain.}}

\shorttitle{S.Toscano \etal Point source searches with the ANTARES.....}
\maketitle

\begin{abstract}
With the installation of its last two lines in May 2008, 
ANTARES is currently the largest neutrino detector in the Northern Hemisphere. The detector comprises 12 detection lines, carrying 884 ten-inch photomultipliers, at a depth of about 2500 m in the Mediterranean Sea, about 40 km off shore Toulon in South France. 
Thanks to its exceptional angular resolution, 
better than 0.3$^\circ$ above 10 TeV, and its favorable location with the 
Galactic Center visible 63\% of time, ANTARES is specially suited for the search 
of astrophysical point sources. Since 2007 ANTARES has been taking data 
in smaller configurations with 5 and 10 lines. 
With only 5 lines it already has been possible to set the most restrictive 
upper limits in the Southern sky. In this contribution we present the search 
of point sources with the 5-line data sample.    
\end{abstract}

 
\section{Introduction}
The ANTARES (Astronomy with a Neutrino Telescope and Abyss environmental RESearch) detector~\cite{ANT09}~\cite{ANT08} has been completed in May 2008, and it is currently the major neutrino telescope in the Northern Hemisphere. The full detector consists on a 3$-$dimensional array of photomultipliers set out in 12 lines deployed in the Mediterranean Sea at a depth of about 2500~m and about 40~km from the south coast of France. The 10$^{\prime\prime}$ photomultipliers~\cite{AMR02} detect the Cherenkov light induced by the relativistic muons
generated in the high energy neutrino interactions with the surrounding material.\\
During the year 2007 ANTARES was operated in a smaller configuration of only 5 lines. Nevertheless, the good angular resolution of the 5 line configuration ($<$~0.5$^\circ$ at 10 TeV) made it possible to start with the physics analysis and perform a search for neutrino sources in the visible sky of ANTARES. 
In this contribution we present the results obtained by an unbinned analysis for the search of point sources with the 5-line data. \\
The description of the data is given in section~\ref{data} while the methodology is explained in section~\ref{method}.
The results are compatible with a background fluctuation and hence, the first upper limits on the cosmic neutrino flux set for ANTARES are given in section~\ref{results}.

\section{Data processing and detector performance}
\label{data}
Data used in the analysis correspond to a sample of data taken from February to December 2007\footnote{To exclude periods with high bio-activity only runs with a baseline~$<$~120 kHz and a burst fraction $<$~40\% have been used.}, in which ANTARES was operated in a smaller configuration of only 5 of its 12 lines. Due to several detector operations including the deployment and connexion of new lines, the actual live-time for the 5-line configuration is 140 days.

\begin{figure}[!h]
	  \centering
    \includegraphics[width=2.5in]{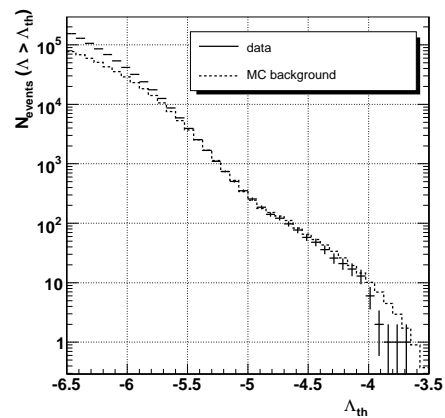}
    \caption{\small{Comparison between data (point) and MC (line) for the complementary cumulative distribution of $\Lambda$. Only up-going tracks with an elevation $< -10^\circ$ are shown.}}
    \label{fig:LAMBDA}
  \end{figure}

The reconstruction of the muon track is achieved by using the information given by the time and amplitude of the signal of the Cherenkov photons arriving to the photomultiplier. The reconstruction algorithm is based on a maximum likelihood method~ \cite{AART} where the quality parameter is defined by the maximum log-likelihood per degree of freedom plus a term that takes into account the number of compatible solutions, N$_{comp}$, found in the algorithm:
\begin{equation}
\Lambda = log(L)/N_{DOF} + 0.1(N_{comp}-1).
\label{eq:lambda}
\end{equation}
\noindent
Fig.\ref{fig:LAMBDA} shows the complementary cumulative distribution\footnote{It is the number of events above any given $\Lambda_{th}$.} of $\Lambda$ for up-going events with a cut in elevation $< -10^\circ$ in order to eliminate a possible atmospheric muon contamination near the horizon. 
	
\noindent

\begin{figure}[!h]
	  \centering
    \includegraphics[width=2.5in]{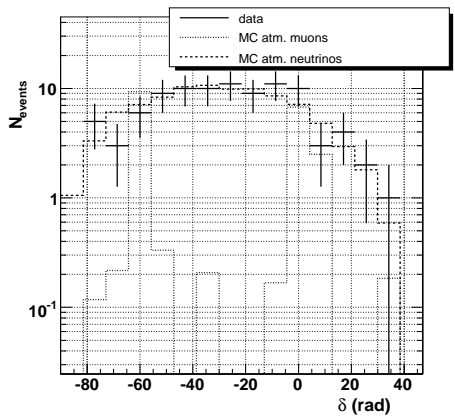}
    \caption{\small{Declination distribution for up-going events. Data (point) and MC (lines) are compared for the quality cuts used in the analysis ($\Lambda > -4.7$, elevation $< -10^\circ$).}}
    \label{fig:DECL}
  \end{figure}

In this analysis, the contributions from atmospheric muons and neutrinos coming from the cosmic ray interactions in the Earth$^\prime$s atmosphere are simulated using Monte Carlo techniques. Atmospheric muons are simulated with CORSIKA~\cite{CORSIKA}, with QGSJET~\cite{QGSJET} models for the hadronic interactions and Horandel~\cite{Horandel} for the cosmic ray composition; neutrinos are generated with the GENNEU~\cite{GENNEU} package and the Bartol~\cite{Bartol} model.\\
In fig.\ref{fig:DECL} the declination distribution for both Monte Carlo and real data is shown. 
After applying a cut in $\Lambda < -4.7$ and elevation$< -10^\circ$ a total number of 94 events are selected with a contamination\footnote{The ratio between the number of atmospheric muons and the total number of events expected from Monte Carlo simulations.} of 20\% due to misreconstructed atmospheric muons.  
This distribution is used in the analysis to estimate the background density function (see next section).

\noindent 
The two plots show a good agreement between data and Monte Carlo.
    
\begin{figure*}[!th]
   \centerline{\subfloat[Effective area]{\includegraphics[width=2.5in]{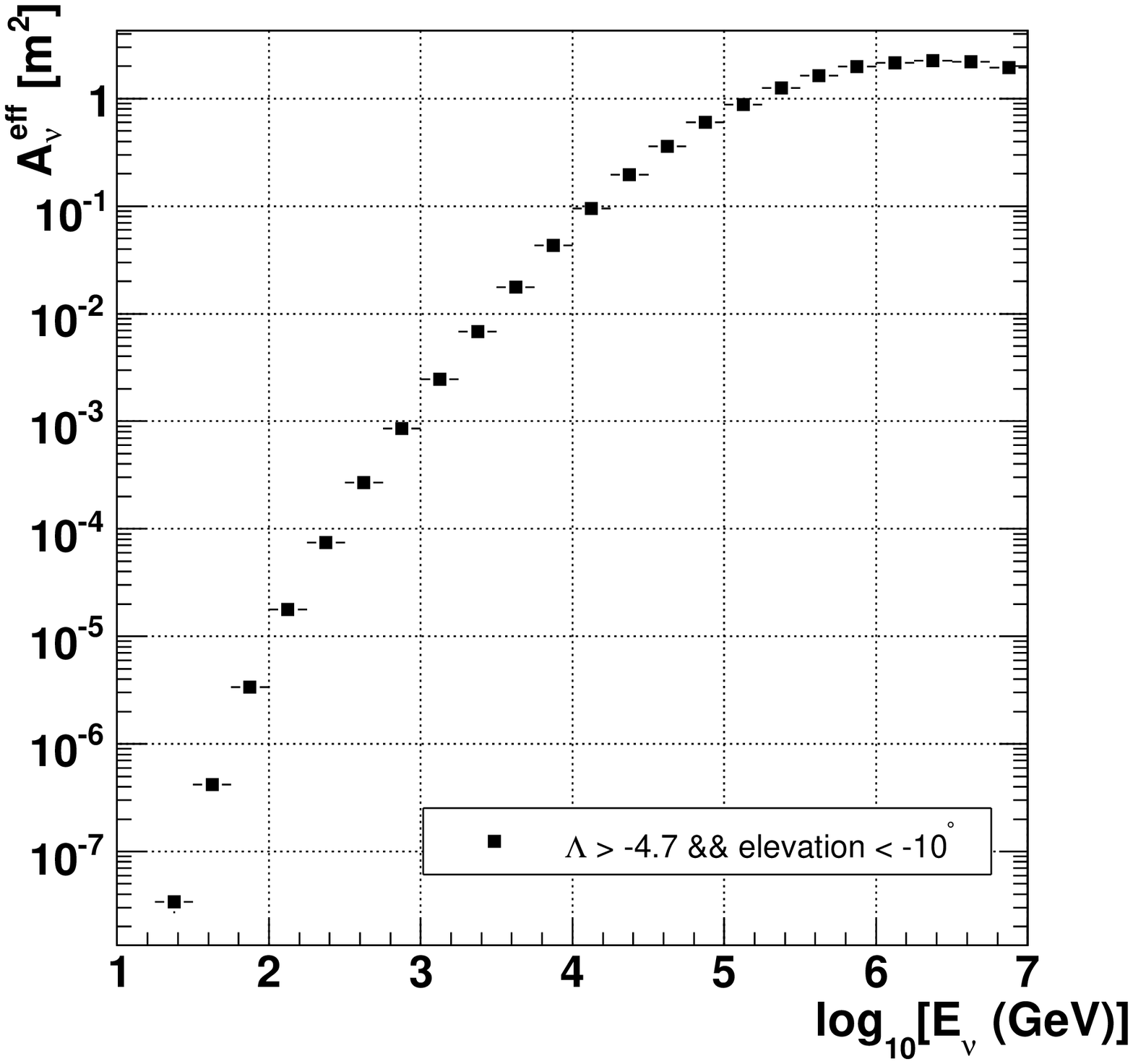} \label{fig:EffA}}
              \hfil
              \subfloat[Angular resolution]{\includegraphics[width=2.5in]{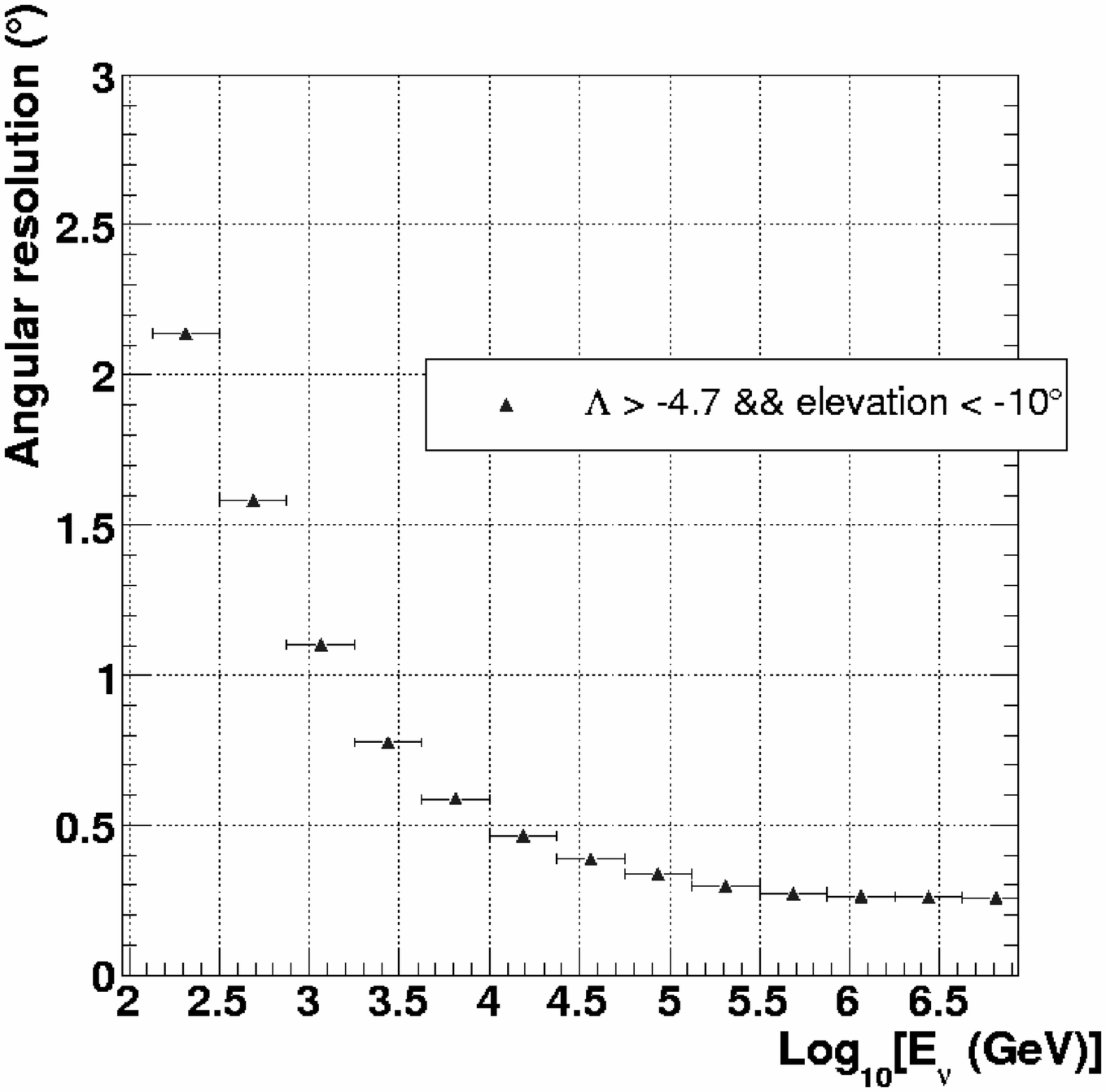} \label{fig:AngRes}}
             }
   \caption{\small{Performance of the ANTARES 5-line detector in terms of effective area (left) and angular resolution (right) as a function of the neutrino energy.}}
   \label{fig:PERFORMANCE}
 \end{figure*}

The detector performance is usually done in terms of its effective area and angular resolution. In fig.\ref{fig:PERFORMANCE} the expected performance for the 5-line detector is shown as a function of the neutrino generated energy.
At 10 TeV the effective area (fig.\ref{fig:EffA}), averaged over the neutrino angle direction, has a value of about $4\times 10^{-2}~$m$^2$. The decrease of the effective area above $\sim$ 1 PeV is due to the opacity of the Earth for neutrinos at these energies. The angular resolution (fig.\ref{fig:AngRes}) for high energy neutrinos is limited by intrinsic detector and environmental characteristics (i.e. PMT transit time spread, dispersion and scattering of light) and it is better than $0.5^\circ$ for energies above 10~TeV. 

\section{Searching algorithms}
\label{method}
Two methods have been developed to perform the search for point sources. 
The first one is an unbinned method based on the expectation-maximization algorithm~\cite{EM} applied to the point source search in a neutrino telescope~\cite{JUANAN}. The second method is a standard binned method only used to cross-check the results obtained with the unbinned method.

\subsection{Unbinned method}

The expectation-maximization method is a clustering algorithm that analytically maximizes the likelihood in finite mixture problems which are described by different density components (pdf). In our case the mixture problem can be expressed as the sum of two density distributions: the background distribution of atmospheric neutrinos and the signal of cosmic neutrinos. 
Since the detector response is uniform in right ascension the background only depends on the declination. A fit of the declination distribution of real data (fig.\ref{fig:DECL}) gives the background density used in the algorithm. The signal pdf model is assumed to follow a two-dimensional Gaussian.
The EM algorithm assumes that the set of given observations (in our case the equatorial coordinates of the observed neutrino events) forms a set of incomplete data vectors. The unknown information is whether or not the observed event belongs to the background or the signal pdf. A class indicator vector or weight is added to each event taking the value 0 for background and 1 for signal. The observed events with this additional associated weight forms the complete data set with a new complete likelihood. The EM procedure consists of two main steps: in the expectation step (E-step) an estimate of the initial parameters ($\Psi^{(0)}$), including the value of the associated weights, is given and the expectation value of the complete log-likelihood is evaluated for this current set of parameters; in the Maximization step (M-step) a new set of parameters $\Psi = {\Psi^{(m)}}$ is found that maximize the complete data log-likelihood. Successive maximizations of the complete likelihood lead to the maximization of the likelihood of the incomplete data set.\\
After the likelihood maximization the model testing theory is used to confirm or reject the existence of a point source. 
In this analysis the Bayesian Information Criterion (BIC) is used as a test statistic:

\setlength{\arraycolsep}{0.0em}
  \begin{eqnarray}
 BIC=2\log p(\{x\}|\Psi_{1}^{ML},M_{1}) \nonumber\\
  - 2\log p(\{x\} | M_{0})- \nu_1 \log(n), 
\label{eq:BIC}
  \end{eqnarray}              %
  \setlength{\arraycolsep}{5pt}


\noindent
which is the maximum likelihood ratio of the two models we are testing ($M_1 =$ background + signal, over $M_0 =$ background) plus a factor that takes into account the degrees of freedom in our testing model ($\nu_1$) and the number of events $n$ in our sample; $\Psi_k^{ML}$ indicates the estimate obtained by the algorithm for the set of parameters that define each model. \\
Since there is no analytical probability distribution associated to our problem $10^4$ experiments have been simulated to infer the BIC distribution. The background is simulating by randomizing real events uniformly in right ascension. The signal simulation has been inferred from Monte Carlo using the corresponding angular error distributions for neutrino spectral index of 2.0. 

In case of an observation a $BIC_{obs}$ value is selected and the probability that this $BIC_{obs}$ is not due to a source is computed. When the probability reaches the 90\% it can be ensured that there is no source emitting a flux that would yield a BIC value higher than the one observed and an upper limit is set.  

\subsection{Binned method}

Since it is well known that binned methods are not as sensitive as unbinned analysis, in this work the binned method has been used only to cross-check the results obtained with the EM method.
The point source search analysis presented here consists in the optimization of the size of the search cone in order to maximize the probability of finding a cluster of events incompatible with background (at a 90\% C.L.). The approach followed in this analysis is the minimization of the Model Rejection Factor (MRF). The MRF is defined as the ratio between the average upper limit, which depends on the expected background inside the bin, and the signal contained in the searching bin. \\
The background is estimated from real data and the fraction of the contained signal is computed using the angular error distribution inferred from Monte Carlo simulations for a neutrino flux with a spectral index of 2.0.
The angular radius that minimizes the MRF is considered the optimum bin size for point source search.

\section{Results}
\label{results}
Two different implementations of a point source analysis have been applied. In the first one a search for a signal excess in pre-defined directions in the sky corresponding to the position of neutrino candidate sources is performed. The second is a full sky survey in which no assumptions about the source position is made.\\

The ANTARES collaboration follows a strict blinding policy. This means that the analyses have been optimized using real data scrambled in right ascension. Once the parameters have been fixed the results are obtained by looking at real data.

\subsection{Fixed source search}
The first analysis performed to search for a point-like signal consists in looking at a list of possible neutrino emitters.
Taking into account the ANTARES visibility (for up-going events) and the lack of statistics at declinations higher than $10^\circ$, 24 sources have been selected among the most promising galactic and extragalactic neutrino source candidates (supernova remnants, BL Lac objects, etc.). The hot spot reported by the IceCube collaboration in the analysis of the 22-line data \cite{IC22} has also been included in the ANTARES list. 
No significant excess has been found in the 5-line data sample. The results are shown in table~\ref{tab:sources}. The P-values, i.e. the probability of the background to produce the observed BIC value, has been reported for each candidate source in the list. The lowest P-value (a 2.8 $\sigma$ excess, pretrial) corresponds to the location ($\delta=$ -57.76$^\circ$, RA=155.8$^\circ$), which is expected in 10\% of the experiments when looking at 25 sources (post-trial probability).
The results are compatible with the binned method.

\begin{figure}[!h]
	  \centering
    \includegraphics[width=3.0in]{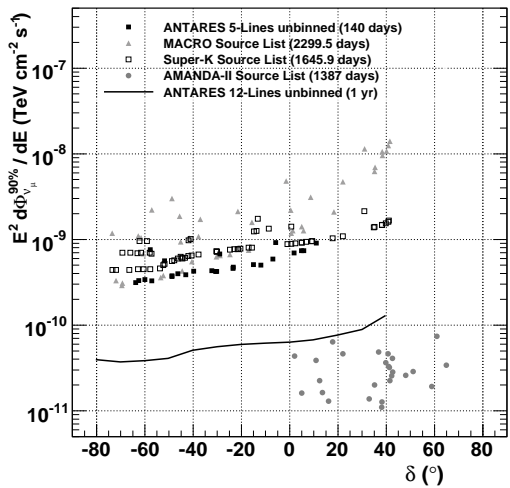}
    \caption{\small{Upper Limits for ANTARES 5-line data (black-filled squares) compared with the results published by other neutrino experiments (MACRO \cite{MACRO}, AMANDA \cite{AMANDA} and SuperKamiokande \cite{SuperK}). The predicted sensitivity of ANTARES for 365 days (line) is also shown.}}
    \label{fig:LIMIT}
  \end{figure}

\begin{table}
 \begin{center}
{\footnotesize
\begin{tabular}{|l|c|c|c|c|} \hline 

Source name &    $\delta~(^\circ)$ &    RA ($^\circ$) &     P-value &    $\phi_{90}$ \\
            &                      &                  &             &                 \\ 
\hline
PSR B1259-63      &    -63.83 &    195.70  &     1  &    3.1  \\
RCW 86            &    -62.48 &    220.68  &     1  &    3.3   \\
ESO 139-G12       &    -59.94 &    264.41  &     1  &    3.4   \\
HESS J1023-575    &    -57.76 &    155.83  &  0.004 &    7.6   \\
CIR X-1           &    -57.17 &    230.17  &     1  &    3.3   \\
HESS J1614-518    &    -51.82 &    243.58  &  0.088 &    5.6   \\
PKS 2005-489      &    -48.82 &    302.37  &     1  &    3.7   \\
GX 339            &    -48.79 &    255.70  &     1  &    3.8   \\
RX J0852.0-4622   &    -46.37 &    133.00  &     1  &    4.0   \\
Centaurus A       &    -43.02 &    201.36  &     1  &    3.9   \\
RX J1713.7-3946   &    -39.75 &    258.25  &     1  &    4.3   \\
PKS 0548-322      &    -32.27 &    87.67   &     1  &    4.3   \\
H 2356-309        &    -30.63 &    359.78  &     1  &    4.2   \\
PKS 2155-304      &    -30.22 &    329.72  &     1  &    4.2   \\
Galactic Centre   &    -29.01 &    266.42  &  0.055 &    6.8   \\
1ES 1101-232      &    -23.49 &    165.91  &     1  &    4.6   \\
W28               &    -23.34 &    270.43  &     1  &    4.8   \\
LS 5039           &    -14.83 &    276.56  &     1  &    5.0   \\
1ES 0347-121      &    -11.99 &    57.35   &     1  &    5.0   \\
HESS J1837-069    &    -6.95  &    279.41  &     1  &    5.9   \\
3C 279            &    -5.79  &    194.05  &   0.030&    9.2   \\
RGB J0152+017     &     1.79  &    28.17   &     1  &    7.0   \\
SS 433            &     4.98  &    287.96  &     1  &    7.3   \\
HESS J0632+057    &     5.81  &    98.24   &     1  &    7.4   \\
IC22 hotspot      &    11.00  &    153.00  &     1  &    9.1   \\ \hline

\end{tabular}
}
 \caption{\small{Flux upper limits for 25 neutrino source candidates. Together with the source name and
 location in equatorial coordinates, the P-value and the 90\% confidence level upper limit for $\nu_{\mu}$ flux with $E^{-2}$ spectrum (i.e. $E^2d\phi_{\nu_{\mu}}/dE~\le~\phi_{90}~\times~10^{-10}~TeV~cm^{-2}~s^{-1}$) over the energy range 10 GeV to 100 TeV are provided.}}
 \label{tab:sources}
 \end{center}
\end{table}
\noindent
The corresponding flux limits have been also reported in the last column and shown in fig.\ref{fig:LIMIT}. As can be seen, the first limits from ANTARES are already competitive with the previous experiments looking at the Southern sky.

\subsection{All sky search}
In the all sky survey no assumption about the source location is done. 
In the EM method a pre-clustering algorithm selects few cluster candidates and only the cluster with the highest significance is considered.  
The highest BIC value is found at a position ($\delta=$ -63.7$^\circ$, RA=243.9$^\circ$), corresponding to a P-value of 0.3 (1$\sigma$ excess). In this case no trial corrections has to be done because the method looks directly to the most significant cluster.  \\

\section{Conclusions}
ANTARES is the first undersea neutrino telescope and the largest apparatus in the Northern Hemisphere. It reached its full configuration in May 2008. Due to its exceptional angular resolution better than 0.3$^\circ$ at high energy it is especially suited for the search for neutrino point sources. In this contribution the analysis of the data taken with 5 lines during the year 2007 have been presented together with the first limits for a point source search. Two independent methods have been used: a more powerful unbinned method based on the Expectation-Maximization algorithm and a binned method to cross-check the results. 
The two methods have been applied to search for a signal excess at the locations of a list of candidate sources (fixed source search). No statistically significant excess have been found in the data. 
The lowest P-value is a 2.8 $\sigma$ excess (pre-trial) corresponding to the location ($\delta=$ -57.76$^\circ$, RA=155.8$^\circ$), expected in 10\% (post-trial) of the experiments when looking at 25 sources.\\
Moreover a \emph{blind} search over the full sky has been also performed, without making any assumption about the source position. Again, no excess has been found and the lowest P-value corresponds to $1\sigma$ excess at ($\delta=$ -63.7$^\circ$, RA=243.9$^\circ$).\\
Although no evidence of neutrino sources have been found in the data, with only 5 lines and 140 days of livetime, ANTARES has set the best upper limits for neutrino point sources in the Southern Hemisphere. \\
The analysis of new data is already underway. The new data sample corresponds to data taken during the year 2008, in which detector was taking data in different configurations (due to the deployment of new lines). The results will be presented soon.

\end{document}